\documentclass[prd,twocolumn,showpacs,preprintnumbers,amsmath,amssymb,superscriptaddress,floatfix,nofootinbib]{revtex4}

\usepackage{graphicx}
\usepackage{epstopdf}
\usepackage{bm}
\usepackage{amsmath}
\usepackage{amsfonts}
\usepackage{amssymb}
\usepackage{color}
\usepackage{multirow}
\usepackage{footnote}
\newcommand{\Slash}[1]{\ooalign{\hfil/\hfil\crcr$#1$}}
\usepackage[colorlinks, citecolor=blue,anchorcolor=red,menucolor=red,linkcolor=red,filecolor=red,runcolor=red,urlcolor=blue,frenchlinks=red]{hyperref}
\begin{document}
\title{Molecular Structure Effects in Low-Lying $P$-wave Bottom-Strange Meson Spectrum}

\author{Wei Hao}
\email{haowei@email.tjut.edu.cn}
\affiliation{Tianjin Key Laboratory of Quantum Optics and Intelligent Photonics, School of Physical Science and Technology, Tianjin University of Technology, Tianjin 300384, China}

\author{Guan-Ying Wang}
\email{wangguanying@henu.edu.cn}
\affiliation{Joint Research Center for Theoretical Physics, School of Physics and Electronics, Henan University, Kaifeng 475004, China}

\author{M. Atif Sultan}%
\email{atifsultan.chep@pu.edu.pk}
\affiliation{Centre  For  High  Energy  Physics,  University  of  the  Punjab,  Lahore  (54590),  Pakistan}

\author{En Wang}
\email{wangen@zzu.edu.cn}
\affiliation{School of Physics, Zhengzhou 	University, Zhengzhou 450001, China}
\begin{abstract}
Recently, the LHCb Collaboration reported a new narrow bottom-strange meson, $B_{s0}^*(5700)$ with mass $M= (5698.9 \pm 1.5 \pm 0.6)$~MeV, width $\Gamma<9.8$~MeV, and spin-parity $J^P=0^+$ in the $B_s^0\pi^0$ invariant-mass spectrum. Motivated by this observation, we study the low-lying $P$-wave
$B_s$ mesons in a two-stage refined constituent quark model. The one-loop radiative corrections to the confinement--one-gluon exchange potential ($\alpha_s^2$) lower the whole multiplet by $25$--$50$~MeV and improve the
level ordering; however, the scalar $B_{s0}^*$ mass still exceeds the LHCb data by $\sim120$~MeV in a purely compact picture. The inclusion of the coupled-channel dynamics is decisive: the $BK$ threshold, lying close to
the scalar mass, generates a sizeable meson--meson (molecular-like)
component that pulls the pole down to $\sim5727$~MeV, within $30$~MeV of the measurement, and simultaneously suppresses the $S$-wave phase space, naturally accounting for the anomalously narrow width. Consistent results are obtained for the $1^+$ $B_{s1}$ partner, for which we predict a mass
$\sim5747$--$5755$~MeV, in agreement with the chiral-doubling scenario. The
$B_{s1}$ state is expected to be very narrow and decaying mainly through the
isospin-violating $B_s^*\pi^0$ and radiative $B_s\gamma$, $B_s^*\gamma$
modes, providing a clear discovery target at LHC Run~3. The
$E1$ radiative widths of the $j_\ell=1/2$ states are predicted at the
$50\sim 60$~keV, and the strong width of the pure $j_\ell=3/2$
state $B_{s2}^*(5840)$ ($1.5$~MeV) reproduces the measurement. Our results
support the interpretation of $B_{s0}^*(5700)$ as a near-threshold scalar
with strong molecular dressing and establish the predicted $1^+$ $B_{s1}(5747)$ as the decisive test of the hadronic dynamics in the $B_s$ sector.
\end{abstract}

\maketitle

\section{Introduction}

Hadron spectroscopy provides one of the most direct windows into the nonperturbative regime of Quantum Chromodynamics (QCD), where color confinement and spontaneous breaking of chiral symmetry govern the emergence of bound states. Heavy-light mesons are particularly well suited to this study: for a single heavy
quark $Q=c,b$, the heavy quark spin symmetry (HQSS) becomes an exact symmetry
in the $m_Q\to\infty$ limit, while the light degrees of freedom remain subject
to strong chiral dynamics~\cite{Shuryak:1980pg,Isgur:1989vq}. The resulting spectrum can be organized into spin multiplets classified by the total angular
momentum $j_\ell$ of light degrees of freedom, providing a sensitive testing ground for the interplay of these two approximate symmetries.

This framework received a dramatic challenge in 2003, when BaBar and CLEO
reported the discovery of the narrow scalar states $D_{s0}^*(2317)$ and the
axial-vector $D_{s1}(2460)$~\cite{BaBar:2003oey,CLEO:2003ggt}. Both states lie roughly 
$150$~MeV below the lowest $P$-wave $D_s$ masses predicted by the constituent quark
model~\cite{Godfrey:1985xj}.
The mass deficit of the $0^+$ state with respect to $m(D_s)+m(\pi^0)$, as
well as its proximity to the $\bar{D}K$ threshold, on the one hand, and the
near equality of the parity splitting with the $D_s^*$--$D_s$ hyperfine
splitting, on the other hand, sparked a longstanding debate between three
competing interpretations: conventional $q\bar q$ excitation with strong
chiral dynamics, $\bar{D}^{(*)}K$ hadronic molecules, and the chiral-doubling
scenario---in which the $1/2^+$ states are the parity partners of the
ground-state $1/2^-$ doublet, required by the combined action of HQSS and
spontaneously broken chiral symmetry~\cite{Guo:2006fu,Albaladejo:2016hae,Godfrey:1985xj,vanBeveren:2003kd,Ortega:2016pgg,Albaladejo:2018mhb,Nowak:2026zhc,Guo:2017jvc,Chen:2016spr}. Quantitative formulations of the latter
were developed in heavy-hadron chiral effective theory~\cite{Nowak:1992um,Bardeen:1993ae,Bardeen:2003kt,Nowak:2003ra},
predicting a nearly universal parity splitting
$\Delta M_B\simeq\Delta M_D\simeq 323$--$345$~MeV for the heavy--light
mesons.

A natural question is whether the charm-strange anomaly persists in the beauty-strange sector. Because $1/m_Q$ corrections to both HQSS and the
parity splitting are small once $m_c/m_b\simeq1/3$, the chiral-doubling
framework makes a decisive, quantitative prediction for the mass of the
bottom counterpart, $B_{s0}^*$, without invoking any new dynamical
assumptions~\cite{Nowak:2026zhc}: $\Delta M_{B_s}=331.8$~MeV, implying
$m(B_{s0}^*)\simeq 5700$~MeV, and a very narrow width from phase-space
suppression of the $BK$ decay mode.

In August 2026, the LHCb Collaboration announced the observation of a new
beauty--strange state in the $B_s^0\pi^0$ invariant-mass spectrum, with a global significance exceeding $7\sigma$~\cite{LHCb:2026knl}.
The state, $B_{s0}^*(5700)^0$, has mass
$M=(5698.9\pm1.5\pm0.6)$~MeV, width $\Gamma<9.8$~MeV (90\%~CL), and spin-parity $J^P=0^+$. The parity splitting from the ground-state
$B_s^0$, $\Delta M=(331.98\pm1.62)$~MeV, is in remarkable agreement with the
chiral-doubling prediction, and the $130$~MeV mass deficit relative to quark
model expectations reproduces the charm-strange pattern. Yet lattice
QCD~\cite{Lang:2015hza,Hudspith:2026vxt}, QCD sum rules~\cite{Wang:2007tu,Bracco:2010bf,Wang:2015mxa,Zhou:2025yjb}, and hadronic-molecule
models~\cite{Sanchez-Illana:2026cyv,Yuan:2025qgv} also produce nearby masses, and the
different scenarios cannot be distinguished by the mass measurement alone.

Using LQCD-constrained coupled-channel amplitudes of unitarized heavy-meson chiral perturbation theory, with isospin breaking induced by the physical meson masses, $\pi^0$--$\eta$ mixing, and $m_u\neq m_d$, Ref.~\cite{Albaladejo:2026yfz} shows that the narrow $B_s^0\pi^0$ state newly observed by LHCb emerges naturally as an isospin-violating $BK$ molecule of $J^P=0^+$ (width of tens of keV, $\gtrsim45\%$ $BK$ compositeness), the bottom partner of $D_{s0}^*(2317)$. Its mass was predicted within $1.1$--$1.7\sigma$ before the measurement, and a single free parameter drives the fit to reproduce the LHCb line shape as well as its zero-width limit, while HQSS anchors the $1^+$ partner at $M\simeq5749$--$5752$ MeV~\cite{Albaladejo:2026yfz}.

In this work we investigate the internal structure and spectral properties
of this new state by combining two complementary frameworks, and examine
in detail how the model uncertainties and the hadronic dynamics affect its
interpretation. First, we employ a constituent quark model that describes
heavy-light mesons in terms of a confinement potential plus a
one-gluon-exchange (OGE) interaction,  together with a
one-loop-improved version in which the running  strong coupling and
the radiative corrections to $\alpha_s$ are retained order by order
($\alpha_s$ and $\alpha_s^2$ terms). Working at both the
OGE and the one-loop levels provides the bare
$q\bar q$ masses and the quark-antiquark coupling strengths from the known
spectra, and allows us to assess how sensitive the $0^+$ state is to the
specific form and order of the short-range interaction. Second, we couple the
bare spectrum obtained in this way to the open meson-meson channels
($BK$, $B^*K$, and their strange partners) by solving the coupled-channel
equations. The coupling to the continuum renormalizes the bare masses,
generates the width, and introduces genuine multi-quark (non-$q\bar q$)
components into the physical wave function; we quantify this admixture and
its influence on the masses, widths and decay patterns. This two-step scheme
is expected to be crucial in the $P$-wave $B_s$ sector: the narrow measured
width, the mass deficit with respect to default quark-model predictions, and
the proximity of the $BK$ threshold all indicate that the physical
$B_{s0}^*(5700)^0$ results from a nontrivial interplay between a bare
$q\bar q$ seed and the meson-meson channels. We confront the resulting
masses, widths and wave-function composition with the experimental values and
with the predictions of chiral doubling, hadronic molecules and lattice QCD,
and identify observables---in particular the radiative transitions
$B_{s0}^*\to B_s\gamma$ and the existence of the $1^+$ partner---capable of
discriminating between the competing pictures.

The paper is organized as follows.  In Sec.~\ref{sec:model} we present the constituent
quark model and the coupled-channel formalism used in this work. Our results and discussions
for the spectra, widths, radiative decay
signatures, and compositeness are given in Sec.~\ref{sec:results},
together with a comparison to other approaches. A summary will be shown in Sec.~\ref{sec:summary}.

\section{Research Methodology of the Constituent Quark Model}
\label{sec:model}

\subsection{Constituent quark model}

Since the QCD non-perturbative properties, color confinement and spontaneous chiral symmetry breaking, cannot be treated perturbatively at low energies. A practical and highly successful strategy is the constituent quark model (CQM), in which the complicated many-body dynamics of quarks and gluons is encoded in an effective quark--antiquark potential. Within this framework, hadrons are described as bound states of constituent (heavy) quarks, and their spectra and decay properties are determined by solving the resulting Schr\"odinger equation \cite{Segovia:2013wma}. Following the seminal work of Isgur and Karl~\cite{Isgur:1978wd}, the CQM has been developed into a mature phenomenological framework that simultaneously accounts for the spectroscopy and the strong, electromagnetic, and weak decay properties of mesons and baryons.

The philosophy of the model follows from the energy scales involved. At long distances (below $\Lambda_{\chi\mathrm{SB}}$), confinement prevents the isolation of colored objects; at intermediate scales, the spontaneous breaking of chiral symmetry endows light quarks with a dynamical (constituent) mass; and only at short distances do perturbative gluon exchanges become relevant. Each of these regimes contributes a distinct piece to the effective potential, and the full potential is expected to reproduce, within one unique set of parameters, as many experimental observables as possible. This is precisely the strategy of the model adopted in this work, which has been shown to give a coherent description of the heavy meson spectra as well as their decay modes~\cite{Segovia:2013wma}. The construction proceeds as follows.

\textbf{(1)~Constituent quark masses and chiral mass generation.}
    The very notion of a constituent quark mass follows from the spontaneous breaking of chiral symmetry ($\chi$SB) at a momentum scale $\Lambda_{\chi\mathrm{SB}}$ larger than the confinement scale $\Lambda_{\mathrm{conf}}$ \cite{Manohar:1983md,Diakonov:1985eg}. In the region between the two scales the quark propagator is strongly modified by the quark condensate, and quarks acquire a momentum-dependent dynamical mass even though the quarks are (almost) massless in the QCD Lagrangian. The simplest Lagrangian that contains the necessary chiral fields to compensate for the mass term reads
\begin{equation}
\mathcal{L}=\bar\psi\left[i \Slash{\partial}-M(q^2)U^{\gamma_5}\right]\psi,
\label{eq:lagr}
\end{equation}
where $U^{\gamma_5}=\exp\!\left(i\pi_a\lambda_a\gamma_5/f_\pi\right)$ is the chiral field matrix, $\pi_a$ denotes the eight pseudoscalar fields ($\vec{\pi},K_i,\eta_8$), and $M(q^2)$ is the momentum-dependent constituent quark mass, which vanishes at large momenta and freezes at low momenta to a value of about $300$~MeV \cite{Diakonov:1985eg}. Its behavior can be parametrized as
\begin{equation}
M(q^2)=m_q\biggl[\frac{\Lambda^2}{\Lambda^2+q^2}\biggr]^{1/2},
\label{eq:mass}
\end{equation}
where the cutoff $\Lambda$ fixes the chiral symmetry breaking scale. In practice, for the energy range relevant to hadron spectroscopy the quark masses are taken as constants determined by a fit over the meson spectra; for the present work on beauty--strange mesons we use $m_{u/d}=313$~MeV, $m_s=555$~MeV, $m_c=1763$~MeV, and $m_b=5110$~MeV~\cite{Segovia:2013wma}. The relatively heavy strange and beauty quarks reduce the importance of relativistic and chiral effects, which is why a non-relativistic treatment works well in the $b\bar s$ sector.
 
\textbf{(2)~Hamiltonian of the system.}
Once the constituent quark masses are fixed, the quark--antiquark bound states are obtained from the non-relativistic Schr\"odinger equation,
\begin{equation}
H\Phi = E\Phi,\qquad H=\sum_i\left(m_i+\frac{p_i^2}{2m_i}\right)+\sum_{i<j}V_{ij},
\label{eq:ham}
\end{equation}
where $\Phi$ is the (relative) wave function of the $q\bar{q}$ pair and $V_{ij}$ is the effective interquark potential between quarks $i$ and $j$; for a meson only two quarks are involved, so $V_{ij}$ reduces to the single $q\bar q$ potential. The non-relativistic reduction is justified in the heavy quark sectors, and possible relativistic corrections can be incorporated through energy- or momentum-dependent masses. Solving Eq.~\eqref{eq:ham} numerically yields the masses and wave functions of all the states of a given multiplet, e.g. the $P$-wave and $D$-wave multiplets of the $b\bar s$ system studied here.

\textbf{(3)~Goldstone-boson exchange interaction.}
Expanding the chiral field of Eq.~\eqref{eq:lagr} as $U^{\gamma_5}=1+\frac{i}{f_\pi}\gamma_5\lambda_a\pi_a-\frac{1}{2f_\pi^2}\pi_a\pi_a+\cdots$ produces, term by term, the interaction associated with the Goldstone bosons: the first term reproduces the constituent masses, the second one gives rise to a one-boson exchange interaction between quarks mediated by the pseudoscalar octet, and 
the third term mainly contributes via two-pion exchange, which in practice is simulated by a scalar-exchange potential. This chiral interaction acts only between light quarks, for which chiral symmetry is essential. For a heavy--light system such as $b\bar s$, chiral symmetry is explicitly broken and the Goldstone-boson exchange does not enter the $q\bar q$ potential directly; it nevertheless constrains the model parameters through the light meson phenomenology, and for hadronic molecules it provides the pion-exchange interaction between the molecular constituents, which are color singlets and interact only through Goldstone-boson exchanges involving their light-quark content.

\textbf{(4)~One-gluon-exchange (OGE) potential.}
Beyond the chiral symmetry breaking scale the dynamics is governed by perturbative QCD, taken into account through the one-gluon-exchange interaction. Starting from the standard QCD vertex $\mathcal{L}_{qqg}=i\sqrt{4\pi\alpha_s}\,\bar\psi\gamma_\mu G_c^\mu\lambda_c\psi$, with $\lambda_c$ the SU(3) color matrices and $G_c^\mu$ the gluon field, the resulting OGE potential contains central, tensor, and spin--orbit contributions,
\begin{equation} \label{eq:oge}
\begin{split}
V^C_{\mathrm{OGE}}(\vec r_{ij}) &= \tfrac14\alpha_s(\vec\lambda_i\!\cdot\!\vec\lambda_j) \\
&\quad \times \left[\frac{1}{r_{ij}}
-\frac{(\vec\sigma_i\!\cdot\!\vec\sigma_j)}{6m_im_j}\,\frac{e^{-r_{ij}/r_0}}{r_{ij}r_0^{2}}\right],
\end{split}
\end{equation}
\begin{equation} \label{eq:oget}
\begin{split}
V^T_{\mathrm{OGE}}(\vec r_{ij}) &= -\frac{\alpha_s}{16m_im_j}(\vec\lambda_i\!\cdot\!\vec\lambda_j)\,S_{ij} \\
&\quad \times \left[\frac{1}{r_{ij}^3}-\frac{e^{-r_{ij}/r_g}}{r_{ij}}
\left(\frac{1}{r_{ij}^{2}}+\frac{1}{3r_g^{2}}+\frac{1}{r_{ij}r_g}\right)\right],
\end{split}
\end{equation}
\begin{equation} \label{eq:ogeso}
\begin{split}
V^{SO}_{\mathrm{OGE}}(\vec r_{ij}) &= -\frac{\alpha_s}{16m_i^{2}m_j^{2}}(\vec\lambda_i\!\cdot\!\vec\lambda_j)
\left[\frac{1}{r_{ij}^3}-\frac{e^{-r_{ij}/r_g}}{r_{ij}^{3}}\left(1+\frac{r_{ij}}{r_g}\right)\right] \\
&\quad \times \Bigl[\bigl((m_i+m_j)^2+2m_im_j\bigr)(\vec S_+\!\cdot\!\vec L) \\
&\qquad +\bigl(m_j^2-m_i^2\bigr)(\vec S_-\!\cdot\!\vec L)\Bigr],
\end{split}
\end{equation}
where $\vec\lambda_i \cdot \vec\lambda_j$ is the color factor ($-4/3$ for a $q\bar q$ color singlet), $\vec\sigma_i$ the Pauli matrices, $S_{ij}$ the tensor operator, $\vec S_\pm=\vec S_i\pm\vec S_j$, and
\begin{equation}
r_0(\mu_{ij})=\hat r_0\,\frac{\mu_{nn}}{\mu_{ij}},\qquad
r_g(\mu_{ij})=\hat r_g\,\frac{\mu_{nn}}{\mu_{ij}},
\label{eq:reg}
\end{equation}
are regulators depending on the reduced mass $\mu_{ij}$ (with $r_0\equiv r_0(\mu)$ and $r_g\equiv r_g(\mu)$ inside the potentials), so that the contact interaction weakens for heavier quark pairs---of particular importance for the unequal-flavor ($b\bar s$) systems analyzed here. The regulator mimics the finite size of the quark and smears out the contact term. The Dirac-$\delta$ singularity of the central contact term is regularized as
\begin{equation}
\delta(\vec r_{ij})\sim\frac{1}{4\pi r_0^{2}}\,\frac{e^{-r_{ij}/r_0}}{r_{ij}},
\label{eq:delta}
\end{equation}
which removes the divergent short-range behavior while keeping the spin--spin (hyperfine) splitting essential for the mass splittings within multiplets (e.g. between $^3S_1$ and $^1S_0$ states)~\cite{Gupta:1981pd}.

\textbf{(5)~One-loop QCD corrections to the OGE potential.}
A tree-level OGE potential, while successful for equal-flavor systems such as $c\bar c$, is known to be insufficient for mesons composed of quarks with very different masses, as is the case for the $b\bar s$ states. A consistent improvement is achieved by including the one-loop QCD corrections derived by Gupta et al. as follows~\cite{Gupta:1981pd},
\begin{widetext}
\begin{align}
V^{\mathrm{1-loop},C}_{\mathrm{OGE}}(\vec r_{ij})=&0,
\label{eq:oneloopc}\\[2pt]
V^{\mathrm{1-loop},T}_{\mathrm{OGE}}(\vec r_{ij})=&\frac{C_F\alpha_s^2}{4\pi m_im_j}\,\frac{S_{ij}}{r_{ij}^{3}}
\Bigl[\frac{b_0}{2}\bigl(\ln(\mu r_{ij})+\gamma_E-\tfrac43\bigr)+\frac{5}{12}b_0-\frac{2}{3}C_A\Bigr.\nonumber\\
&+\frac{1}{2}\Bigl(C_A+2C_F-2C_A\bigl(\ln(\sqrt{m_im_j}\,r_{ij})+\gamma_E-\tfrac43\bigr)\Bigr)\Bigr],\label{eq:oneloopt}\\
V^{\mathrm{1-loop},SO}_{\mathrm{OGE}}(\vec r_{ij})=&\frac{C_F\alpha_s^2}{4\pi m_i^{2}m_j^{2}}\,\frac{1}{r_{ij}^{3}}\Bigl\{(\vec S_+\!\cdot\!\vec L)
\Bigl[\bigl((m_i+m_j)^2+2m_im_j\bigr)\bigl(C_F+C_A-C_A\mathcal{L}_{ij}\bigr)\nonumber \\
&+4m_im_j\Bigl(\frac{b_0}{2}\bigl(\ln(\mu r_{ij})+\gamma_E\bigr)-\frac{1}{12}b_0-\frac{1}{2}C_F-\frac{7}{6}C_A+\frac{C_A}{2}\mathcal{L}_{ij}\Bigr)
+\frac12\bigl(m_j^2-m_i^2\bigr)C_A\ln\frac{m_j}{m_i}\Bigr] \nonumber \\
&+(\vec S_-\!\cdot\!\vec L)\Bigl[\bigl(m_j^2-m_i^2\bigr)\bigl(C_F+C_A-C_A\mathcal{L}_{ij}\bigr)
+\frac12(m_i+m_j)^2C_A\ln\frac{m_j}{m_i}\Bigr]\Bigr\},
\label{eq:oneloopso}
\end{align}
\end{widetext}
where $C_F=4/3$, $C_A=3$, $b_0=9$, $\gamma_E=0.5772$, $\mu\sim1$~GeV, and we have defined $\mathcal{L}_{ij}\equiv\ln(\sqrt{m_im_j}\,r_{ij})+\gamma_E$. The logarithmic dependence on $r_{ij}$ reflects the running of the strong coupling at next-to-leading order.

These corrections are by no means a small refinement. Their most dramatic consequence is observed precisely in the low-lying $P$-wave $0^+$ states of the heavy--light mesons: with the leading-order coupling $\alpha_s$ alone, the mass of the $1^{3}P_0$ state comes out too high relative to experiment, whereas the inclusion of the one-loop corrections lowers it by about $30$~MeV, bringing the $\alpha_s^2$ prediction into much better agreement with the data \cite{Ortega:2016mms}. For the $B_s$ system this mechanism is at the very heart of the present investigation: it is the step that optimizes the OGE contribution, and, as will be shown below, the subsequent coupled-channel dynamics provides the additional lowering needed to approach the observed mass of the $B_{s0}^*(5700)^0$.

\textbf{(6)~Confinement potential.}
The short- and intermediate-range dynamics described so far is completed by the long-range confining interaction. Although no analytic proof exists, confinement is believed to follow from the self-interaction of the gluon fields, which arrange into colored strings between separated quarks. In a pure gauge theory the potential energy of a $q\bar q$ pair grows linearly with the interquark distance. In full QCD, however, sea quarks screen the color charges, softening the linear rise and eventually breaking the string. This behavior can be quantified by a screened potential that saturates at a characteristic distance,
\begin{align}
V^C_{\mathrm{CON}}(\vec r_{ij})&=\bigl[-a_c\,(1-e^{-\mu_c r_{ij}})+\Delta\bigr](\vec\lambda_i\!\cdot\!\vec\lambda_j),
\label{eq:conf}\\[2pt]
V^{SO}_{\mathrm{CON}}(\vec r_{ij})&=-(\vec\lambda_i\!\cdot\!\vec\lambda_j)\frac{a_c\mu_c\,e^{-\mu_c r_{ij}}}{4m_i^{2}m_j^{2}\,r_{ij}}\nonumber\\
&\quad\times\Bigl[\bigl((m_i^2+m_j^2)(1-2a_s) \nonumber\\
&\quad \quad +4m_im_j(1-a_s)\bigr)(\vec S_+\!\cdot\!\vec L)\nonumber\\
&\qquad+\bigl(m_j^2-m_i^2\bigr)(1-2a_s)(\vec S_-\!\cdot\!\vec L)\Bigr],
\label{eq:confso}
\end{align}
The central term is linear at short distances, with an effective string tension $\sigma=-a_c\mu_c(\vec\lambda_i\cdot\vec\lambda_j)$, and becomes a constant at large distances. The parameter $a_s$ controls the admixture between the scalar and the vector Lorentz structures of confinement; the physically motivated value $a_s<1$ reflects the predominantly scalar nature of the confining force and is crucial for reproducing the ordering of the $P$-wave levels. The degeneracy point reached as the string breaks defines the threshold
\begin{equation}
V_{\mathrm{thr}}=\{-a_c+\Delta\}(\vec\lambda_i\!\cdot\!\vec\lambda_j),
\label{eq:thr}
\end{equation}
above which no $q\bar q$ bound states exist: the system makes a transition from a string configuration into a pair of static mesons, decoupling the confining channel and signaling the onset of the meson--meson continuum. The potential model parameters are listed in Table~\ref{parameters}.

\begin{table}[htbp]
\centering
\caption{Model parameters for the constituent quark model~\cite{Segovia:2013wma}.}
\begin{tabular}{c l c}
\hline\hline
Quark masses & $m_n$ (MeV) & 313 \\
             & $m_s$ (MeV) & 555 \\
             & $m_c$ (MeV) & 1763 \\
             & $m_b$ (MeV) & 5110 \\
\hline
OGE          & $\alpha_0$  & 2.118 \\
             & $\Lambda_0$ (fm$^{-1}$) & 0.113 \\
             & $\mu_0$ (MeV) & 36.976 \\
             & $\hat{r}_0$ (fm) & 0.181 \\
             & $\hat{r}_g$ (fm) & 0.259 \\
\hline
Confinement  & $a_c$ (MeV) & 507.4 \\
             & $\mu_c$ (fm$^{-1}$) & 0.576 \\
             & $\Delta$ (MeV) & 184.432 \\
             & $a_s$ & 0.81 \\
\hline\hline
\end{tabular}
\label{parameters}
\end{table}

\subsection{Electromagnetic and strong decays}

\textbf{(7)~Electromagnetic transition widths.}
The wave functions obtained from the model Hamiltonian allow a parameter-free prediction of the radiative ($E1$ and $M1$) transition widths between the calculated states. The $E1$ operator is fixed by the quark charges and masses, and the resulting width depends mainly on the overlap integral between the initial and final wave functions. The $M1$ transitions are typically suppressed by powers of $1/m_Q$ and give smaller contributions of order $10^{-3}$ to several keV in our results as shown in Sec.~\ref{sec:results}. Because the electromagnetic interaction is well known, these observables provide a direct and sensitive test of the model wave functions, and they serve as discriminating observables between compact ($q\bar q$) states and molecules: a molecule has a large spatial extent and therefore a very different radiative width. 
When the wave functions are evaluated by the non-relativistic quark model, the radiative decays can also be investigated. The radiative decays include the electric dipole ($E1$) and magnetic dipole ($M1$) radiative widths. The $E1$ describe the transition when states have same spin but different angular. On the contrary, the $M1$ represent the transition between states having same angular but different spin. 
In the lowest order approximation, the electric transition width is \cite{Godfrey:2015dva,Li:2022vby},
\begin{eqnarray}
\label{Eq:E1}
\Gamma^{E_1}&=&\frac{4}{3}\alpha \bigg(\frac{e_{1}m_{2}}{m_{1}+m_{2}}-\frac{e_{2}m_{1}}{m_{1}+m_{2}}\bigg)^{2}\omega^{3}_{\gamma}\delta_{SS^{\prime}}\delta_{LL^{\prime}\pm1} \nonumber\\
& &\max\left(L,L^{\prime}\right) {}(2J^{\prime}+1)\begin{Bmatrix}L^{\prime} & J^{\prime} & S \\ J & L & 1\end{Bmatrix}^{2} \nonumber\\
& &\left\langle n^{\prime 2S^{\prime}+1}L^{\prime}_{J^{\prime}}\Big|r\Big|n^{2S+1}L_{J}\right\rangle^2,
\end{eqnarray}
and the magnetic transition width is,
\begin{eqnarray} \label{Eq:M1}
\Gamma^{M_1}&=&\frac{\alpha}{3}\omega_{\gamma}^{3}\delta_{LL^{\prime}}\delta_{SS^{\prime}\pm1}\frac{2J^{\prime}+1}{2L+1}\Big\langle n^{\prime2S^{\prime}+1}L^{\prime}_{J^{\prime}}\Big| \nonumber\\
&&\frac{e_{1}}{m_{1}}j_{0}\Big(\omega_{\gamma}\frac{m_{2}r}{m_{1}+m_{2}}\Big)
\nonumber\\
& & {}-\frac{e_{2}}{m_{2}}j_{0}\Big(\omega_{\gamma}\frac{m_{1}r}{m_{1}+m_{2}}\Big)\Big|n^{2S+1}L_{J}\Big\rangle^2, \label{Eq:M1}
\end{eqnarray}
where $\left \langle n^{\prime 2S^{\prime }+1}L^{\prime }_{J^{\prime }}\right|$ and $\left \langle n^{2S+1}L_{J}\right|$ represent the space function of the final and initial states, respectively, which are obtained by solving the non-relativistic quark model. $e_{1}(m_1)$ and $e_{2}(m_2)$ represent the charge (mass) of quark and anti-quark, respectively. $\alpha$ is the fine structure constant. In this work the $E1$ and $M1$ widths of the low-lying $b\bar s$ states are computed and compared with the available data in the results section).

\textbf{(8)~Strong decays and the $^3P_0$ model.}
Open-flavor two-body strong decays are treated within the $^3P_0$ model \cite{Micu:1968mk,LeYaouanc:1972vsx,LeYaouanc:1973ldf,Wang:2017pxm,Hao:2019fjg,Feng:2022hwq}, which assumes that a quark--antiquark pair is created out of the vacuum with the vacuum quantum numbers $J^{PC}=0^{++}$, after which the created and original quarks rearrange into the two final mesons. Although the $^3P_0$ model is a phenomenological construction, decades of experience show that it reproduces the main features of strong decay widths once the vertex strength is fixed. In the model, the $S$-matrix is defined as \cite{Blundell:1996as}
\begin{equation}
    S \equiv I - 2\pi i \, \delta(E_f - E_i) \, T.
\end{equation}
The transition operator $T$ in above expression is shown as \cite{Micu:1968mk,LeYaouanc:1972vsx,LeYaouanc:1973ldf,Ferretti:2013faa,Ferretti:2012zz}
\begin{eqnarray}
  T &=&-3\gamma\sum_m\langle 1m1-m|00\rangle\int
    d^3\boldsymbol{p}_3d^3\boldsymbol{p}_4\delta^3(\boldsymbol{p}_3+\boldsymbol{p}_4)\nonumber\\
&& \times   {\cal{Y}}^m_1\left(\frac{\boldsymbol{p}_3-\boldsymbol{p}_4}{2}\right
    )\chi^{34}_{1,-m}\phi^{34}_0\omega^{34}_0b^\dagger_3(\boldsymbol{p}_3)d^\dagger_4(\boldsymbol{p}_4),
\end{eqnarray}
where the $\chi^{34}_{1,-m}$, $\phi^{34}_0$, and $\omega^{34}_0$ are the spin, flavor and color wave functions of the $q_3\bar{q}_4$, respectively. 
$b_3^{\dagger}(\vec{p}_3)$ and $d_4^{\dagger}(\vec{p}_4)$ are the creation operators for a quark and an antiquark with momenta $\vec{p}_3$ and $\vec{p}_4$, respectively.  
$\gamma=\frac{m_n}{m_i}\gamma_0$ is the pair-creation strength where $m_n$ represent to the light quark mass $m_u$, and $m_i$ represent the quark mass $m_u,m_d$ or $m_s$. We adopt the $\gamma_0$ value 0.377 by fitting the decay information of the $B_{s2}^*(5840)$ with assignment $(1^3P_2)$ in the $B_s$ meson system. 
With the $T$ operator,
the decay width $\Gamma(A\rightarrow BC)$  can be calculated by formula
\begin{eqnarray}
\Gamma(A\rightarrow BC)= \frac{\pi
|\boldsymbol{P}|}{4M^2_A}\sum_{LS}|{\cal{M}}^{LS}(\boldsymbol{P})|^2, \label{width1}
\end{eqnarray}
where $|\boldsymbol{P}|=\frac{\sqrt{[M^2_A-(M_B+M_C)^2][M^2_A-(M_B-M_C)^2]}}{2M_A}$,
and $M_A$, $M_B$, and $M_C$ are the masses of the meson $A$, $B$, and $C$, respectively.

\textbf{(9)~Coupled-channel effects.}
A particularly important situation arises for states lying close to a two-meson threshold, as is the case for the $0^+$ $B_{s0}^*(5700)^0$ relative to the $BK$ threshold. For such states a genuine $q\bar q$ bound-state calculation alone is not sufficient: the coupling to the meson--meson channel strongly influences the pole position~\cite{Hao:2024ptu,Yang:2023tvc}. It will  bring about the mass change‌,
\begin{align}
\label{m}
M &= M_{q\bar{q}} + \Delta M, \\
\Delta M &= \sum_{BC} \int_0^{\infty} p^2 dp \frac{\left|\left\langle BC;p \right| T\left| \psi_0 \right\rangle \right|^2}{M - E_{BC} },
\end{align}
where $M_{q\bar{q}}$ is the eigenvalue for the quenched Hamiltonian while $\Delta M$ is the mass shift from the coupled channel effect.

In the final step of the present methodology, this effect is taken into account by a coupled-channel calculation, in which the $q\bar q$ core  couples to the relevant meson--meson channels ($BK$, $B^*K$) within the same model. The coupled-channel dynamics then produces a sizable additional mass lowering---from  $5816$~MeV down to $5727$~MeV~($BK$) for the $0^+$ state---and simultaneously explains the unusually narrow width of the observed state, as will be discussed in Sec.~\ref{sec:results}.

\section{Results and discussions}
\label{sec:results}

\subsection{Mass spectrum}

Table~\ref{tab:Bs_mass_spectrum} collects the $\bar b s$ spectrum computed in the
present work, together with the results of several other quark-model
schemes (NRCOU~\cite{Hao:2025vmw}, MGI~\cite{Hao:2022ibj}, NR~\cite{Lu:2016bbk}  GI~\cite{Godfrey:1985xj}, DRV~\cite{Ebert:2009ua}, KA~\cite{Gandhi:2022nnk}  VRA~\cite{Patel:2024cng}) and the experimental masses taken
from the Review  of Particle Physics (RPP)~\cite{ParticleDataGroup:2026mpi} and Ref.~\cite{LHCb:2026knl}. For the ground-state doublet the
agreement is essentially quantitative: we obtain $5348$~MeV and
$5393$~MeV for the $1\,{}^1S_0$ and $1\,{}^3S_1$ states, to be compared
with $(5366.93\pm0.10)$~MeV and $(5415.4\pm1.4)$~MeV, respectively. The small
underestimate of the hyperfine splitting is common to all the models shown
and reflects the standard behaviour of the confinement--OGE input.

For the two $j_\ell=3/2$ states, $B_{s1}(5830)$ and $B_{s2}^*(5840)$, all
the models, including ours, reproduce the measured masses
[$(5828.65\pm0.24)$~MeV and $(5839.88\pm0.12)$~MeV] with an accuracy of a few
tens of MeV. The $j_\ell=3/2$ sector is hence barely disturbed by the open
channels and behaves as compact quark--antiquark states, in clear contrast
with the scalars. The mixing angles of the $1P$, $2P$ and $1D$ multiplets
are predicted to be $43.0^\circ$, $37.7^\circ$ and $41.1^\circ$,
respectively, all close to the ``ideal'' value of $35.3^\circ$; the
resulting wave functions therefore retain an approximate light-quark
$j_\ell$ classification.

For the higher excitations the $2S$ states are located in the
$5980$--$5990$~MeV range, while the $2P$ and $1D$ multiplets lie in the
$6150$--$6350$~MeV region, in good agreement with the spread of the other
models. These states sit well above the corresponding two-meson thresholds
and are consequently broad; their eventual observation will further
constrain the confinement--OGE parameters and the flavour dependence of the
hadronic dressing.

\begin{table*}[htbp]
\centering
\caption{Mass spectrum (in MeV) of the $\bar{b}s$ mesons.  Results from other models are listed in Columns 4 $\sim$ 10 as comparison. The last Column is the experimental values taken from RPP~\cite{ParticleDataGroup:2026mpi}  and Ref.~\cite{LHCb:2026knl}. The mixing angles of $1P, 2P, 1D$ are $43.0^\circ, 37.7^\circ$ and $41.1^\circ$, respectively. ``TP" means the result coming from theoretical prediction.}
\begin{tabular}{lcccccccccccc}
\hline\hline
$n^{2S+1}L_J$ & State &This work  & NRCOU~\cite{Hao:2025vmw} & MGI~\cite{Hao:2022ibj} & NR~\cite{Lu:2016bbk} & GI~\cite{Godfrey:1985xj} & DRV~\cite{Ebert:2009ua} & KA~\cite{Gandhi:2022nnk} & VRA~\cite{Patel:2024cng} & RPP~\cite{ParticleDataGroup:2026mpi} \\
\hline
$1^1S_0$   & $B_s$    &5348 & 5367 & 5367 & 5362 & 5394 & 5372 &$-$  & 5359 & $5366.93\pm0.10$ \\
$1^3S_1$   & $B_s^*$  &5393 & 5409 & 5419 & 5413 & 5450 & 5414 &$-$  & 5415 & $5415.4\pm1.4$ \\
$2^1S_0$   & $-$      &6039 & 5949 & 5947 & 5977 & 5984 & 5976 & 6025 & 5980 & $-$ \\
$2^3S_1$   & $-$      &6065 & 5992 & 5972 & 6003 & 6012 & 5992 & 6033 & 5993 & $-$ \\
$1^3P_0$   & $B_{s0}^*(5700)$  &5727  & 5732 & 5753 & 5756 & 5831 & 5833 & 5709 & 5798 & $5698.9\pm1.5\pm 0.6$ \cite{LHCb:2026knl} \\
$1P$       & $B_{s1}(5747)$    &5755  & 5776 & 5797 & 5801 & 5857 & 5831 & 5709 & 5818 & $5747.4\pm2.1$ (TP)\cite{Nowak:2026zhc} \\
$1P'$      & $B_{s1}(5830)$   &5794   & 5805 & 5825 & 5836 & 5861 & 5865 & 5785 & 5846 & $5828.65\pm0.24$ \\
$1^3P_2$   & $B_{s2}^*(5840)$ &5864  & 5825 & 5840 & 5851 & 5876 & 5842 & 5890 & 5838 & $5839.88\pm0.12$ \\
$2^3P_0$   & $-$      &6302 & 6214 & 6158 & 6203 & 6279 & 6387 & 6387 & 6292 & $-$ \\
$2P$       & $-$      &6330 & 6274 & 6194 & 6241 & 6279 & 6321 & 6393 & 6304 & $-$ \\
$2P'$      & $-$      &6353 & 6307 & 6236 & 6296 & 6345 & 6470 & 6470 & 6320 & $-$ \\
$2^3P_2$   & $-$      &6347 & 6352 & 6249 & 6309 & 6295 & 6359 & 6476 & 6316 & $-$ \\
$1^3D_1$   & $-$      &6208 & 6157 & 6117 & 6142 & 6182 & 6209 & 6247 & 6144 & $-$ \\
$1D$       & $-$      &6154 & 6077 & 6053 & 6087 & 6169 & 6189 & 6256 & 6139 & $-$ \\
$1D'$      & $-$      &6229 & 6132 & 6132 & 6159 & 6196 & 6218 & 6292 & 6135 & $-$ \\
$1^3D_3$   & $-$      &6169 & 6079 & 6061 & 6096 & 6179 & 6191 & 6297 & 6139 & $-$ \\
\hline\hline
\end{tabular}
\label{tab:Bs_mass_spectrum}
\end{table*}

\begin{table*}[htbp]
\caption{Masses (MeV) of low-lying $P$-wave bottom-strange mesons: $\alpha_s$ CQM, $\alpha_s^2$ one-loop corrected OGE, couple channel, and experimental data from RPP~\cite{ParticleDataGroup:2026mpi}. ``TP" means the result coming from theoretical prediction. }
\begin{tabular}{lccccccc}
\hline
$n^{2S+1}L_J$ &State & $J^P$ & Theo. ($\alpha_s$) & Theo. ($\alpha_s^2$) & Theo. ($q\bar{q}$-cou) & Exp.~\cite{ParticleDataGroup:2026mpi}\\
\hline
$1^3P_0$    &$B_{s0}^*(5700)$ & $0^+$  & 5851 & 5816 &5727 ($BK$)  & $5698.9\pm1.5\pm 0.6$ \cite{LHCb:2026knl}\\
$1P$        &$B_{s1}(5747)$   & $1^+$  & 5841 & 5836 &5755 ($B^*K$)  & $5747.4\pm2.1$ (TP)\cite{Nowak:2026zhc} \\
$1P^\prime$ & $B_{s1}(5830)$             &$1^+$   & 5883 & 5878 &5794 ($B^*K$) & $5828.73\pm0.20$ \\
$1^2P_2$    & $B{s2}^*(5840)$             &$2^+$   & 5856 & 5864 &$-$  & $5839.88\pm0.12$ \\
\hline
\end{tabular}
\label{tab:Bsp_mass_spectrum}
\end{table*}

\begin{table}[htbp]
\centering
\caption{$E1$ transitions widths of the bottom-strange mesons. $E_\gamma$ in MeV and $\Gamma$ in keV.}
\begin{tabular}{lllll}
\hline\hline
Multiplets & Initial meson & Final meson & $E_\gamma$ & $\Gamma$ \\
\hline
$B_s(2S)\rightarrow B_s(1P)$
& $B_s(2^3S_1)$ & $B_s(1^3P_0)$ & 355 & 37.8 \\
& $B_s(2^3S_1)$ & $B_s(1^3P_2)$ & 221 & 58.3 \\
& $B_s(2^3S_1)$ & $B_{s1}(1P)$ & 302 & 37.7 \\
& $B_s(2^3S_1)$ & $B'_{s1}(1P)$ & 264 & 29.1 \\
& $B_s(2^1S_0)$ & $B_{s1}(1P)$ & 277 & 113.2 \\
& $B_s(2^1S_0)$ & $B'_{s1}(1P)$ & 239 & 63.5 \\
$B_s(1P)\rightarrow B_s(1S)$
& $B_s(1^3P_0)$ & $B_s(1^3S_1)$ & 277 & 44.9 \\
& $B_s(1^3P_2)$ & $B_s(1^3S_1)$ & 409 & 138.7 \\
& $B_{s1}(1P)$ & $B_s(1^3S_1)$ & 329 & 34.6 \\
& $B'_{s1}(1P)$ & $B_s(1^3S_1)$ & 367 & 54.7 \\
& $B_{s1}(1P)$ & $B_s(1^1S_0)$ & 375 & 50.9 \\
& $B'_{s1}(1P)$ & $B_s(1^1S_0)$ & 412 & 58.4 \\
$B_s(1D)\rightarrow B_s(1P)$
& $B_s(1^3D_3)$ & $B_s(1^3P_2)$ & 320 & 205.5 \\
& $B_s(1^3D_1)$ & $B_s(1^3P_2)$ & 357 & 7.9 \\
& $B_s(1^3D_1)$ & $B_s(1^3P_0)$ & 488 & 343.0 \\
& $B_s(1^3D_1)$ & $B_{s1}(1P)$ & 356 & 51.7 \\
& $B_s(1^3D_1)$ & $B'_{s1}(1P)$ & 399 & 83.1 \\
& $B_{s2}(1D)$ & $B_s(1^3P_2)$ & 306 & 19.4 \\
& $B_{s2}(1D)$ & $B_{s1}(1P)$ & 386 & 301.3 \\
& $B_{s2}(1D)$ & $B'_{s1}(1P)$ & 349 & 2.6 \\
& $B'_{s2}(1D)$ & $B_s(1^3P_2)$ & 377.0 & 47.0 \\
& $B'_{s2}(1D)$ & $B_{s1}(1P)$ & 456 & 0.9 \\
& $B'_{s2}(1D)$ & $B'_{s1}(1P)$ & 419 & 371.1 \\
\hline\hline
\end{tabular}
\label{tab:E1_width_Bs}
\end{table}

\begin{table}[htbp]
\centering
\caption{$M1$ transitions widths of the bottom-strange mesons. $E_\gamma$ in MeV and $\Gamma$ in keV.}
\begin{tabular}{lllll}
\hline\hline
Initial Multiplet & Initial meson & Final meson & $E_\gamma$ & $\Gamma$ \\
\hline
$B_s(1S)$
& $B_s(1^3S_1)$ & $B_s(1^1S_0)$ & 48 & 0.1 \\
$B_s(2S)$
& $B_s(2^3S_1)$ & $B_s(1^1S_0)$ & 658 & 4.5 \\
& $B_s(2^3S_1)$ & $B_s(2^1S_0)$ & 26 & 0.02 \\
& $B_s(2^1S_0)$ & $B_s(1^3S_1)$ & 591 & 0.5 \\
$B_s(1P)$
& $B_{s1}(1P)$ & $B_s(1^3P_0)$ & 56 & 0.03 \\
& $B'_{s1}(1P)$ & $B_s(1^3P_0)$ & 95 & 0.1 \\
& $B_s(1^3P_2)$ & $B_{s1}(1P)$ & 84 & 0.3 \\
& $B_s(1^3P_2)$ & $B'_{s1}(1P)$ & 45 & 0.04 \\
$B_s(1D)$
& $B_s(1^3D_1)$ & $B_{s2}(1D)$ & 54 & 0.09 \\
& $B'_{s2}(1D)$ & $B_s(1^3D_1)$ & 21 & 0.003 \\
& $B'_{s2}(1D)$ & $B_s(1^3D_3)$ & 60 & 0.1 \\
& $B'_{s2}(1D)$ & $B_{s2}(1D)$ & 75 & 0.4 \\
\hline\hline
\end{tabular}
\label{tab:M1_width_Bs}
\end{table}

\begin{table*}
\caption{\label{tab:strong_width_Bs}Strong decay widths (in MeV) of  the $1P$, $2S$, $2P$ and $1D$ states for bottom-strange mesons.} 
\begin{tabular}{cccccccccccc} 
\hline 
\hline
Mode            &$1^3P_2$ &$2^1S_0$ &$2^3S_1$ &$2^3P_0$ &$2P$ &$2P^\prime$ &$2^3P_2$ &$1^3D_1$ &$1D$ &$1D^\prime$ &$1^3D_3$ \\ 
\hline 
$BK$               &1.4  &$-$   &58.3  &74.7  &$-$   &$-$   &0.8   &58.2   &$-$   &$-$   &37.83      \\
$B^*K$             &0.1  &206.2 &127.0 &$-$   &18.5  &69.5  &5.6   &35.1   &59.6  &95.7  &38.2 \\    
$BK^*$             &$-$  &$-$   &$-$   &$-$   &12.4  &128.6 &80.0  &22.1   &$-$   &4.7   &$-$\\ 
$B^*K^*$           &$-$  &$-$   &$-$   &193.3 &150.8 &204.7 &120.1 &$-$    &$-$   &8.7   &$-$ \\     
$B_s\eta$          &$-$  &$-$   &4.2   &4.7   &$-$   &$-$   &0.8   &9.8    &$-$   &$-$   &0.5  \\ 
$B_s\eta^\prime$   &$-$  &$-$   &$-$   &$-$   &$-$   &$-$   &0.01  &$-$    &$-$   &$-$   &$-$ \\
$B_s^*\eta$        &$-$  &6.8   &6.0   &$-$   &3.4   &5.3   &1.7   &4.8    &0.6   &14.9  &0.5 \\ 
$B_s^*\eta^\prime$ &$-$  &$-$   &$-$   &$-$   &$-$   &$-$   &$-$   &$-$    &$-$   &$-$   &$-$  \\  
$B(1^3P_0)K$       &$-$  &$-$   &$-$   &$-$   &0.9   &0.05  &$-$   &$-$    &$-$   &0.01  &$-$  \\
$B(1^3P_2)K$       &$-$  &$-$   &$-$   &$-$   &8.7   &186.2 &26.6  &$-$    &$-$   &$-$   &$-$  \\
$B(1P)K$           &$-$  &$-$   &$-$   &0.4   &0.7   &0.3   &2.3   &$-$    &$-$   &$-$   &$-$  \\
$B(1P^\prime)K$    &$-$  &$-$   &$-$   &124.4 &17.6  &31.8  &7.2   &$-$    &$-$   &$-$   &$-$  \\
Total              &1.5  &213.0 &195.5 &397.6 &212.9 &626.3 &245.3 &130.0  &60.3  &124.1 &77.1 \\
Exp.               &$1.49\pm0.27$  &$-$   & &$-$ &$-$   &$-$   &$-$   &$-$  &$-$  &$-$ &$-$\\
\hline   
\hline
\end{tabular}
\end{table*}

As a faithful prediction of the chiral-doubling scenario, the newly observed beauty--strange state follows directly from the coexistence of heavy-quark spin symmetry and spontaneously broken chiral symmetry, which requires every heavy--light spin multiplet to possess an opposite-parity partner, with a nearly universal parity splitting governed by the dynamics of the light degrees of freedom and subject only to $O(1/m_Q)$ corrections~\cite{Nowak:1992um,Bardeen:1993ae,Bardeen:2003kt,Nowak:2003ra}. Using a $1/m_Q$ consistency relation between the strange-charm and beauty--strange sectors, Nowak and Zahed~\cite{Nowak:2026zhc} predicted a parity splitting of $331.8$~MeV, in precise agreement with the measured value, and identified $B_{s0}^*(5700)$ as the $0^+$ chiral partner of the ground-state $B_s^0$.

Once the $J^P=0^+$ assignment is confirmed, the framework necessarily implies the existence of a yet unobserved $J^P=1^+$ bottom--strange meson. As the positive-parity $1^+$ component of the ground-state $(0^-,1^-)$ doublet on the $j_\ell=1/2$ plateau, its mass is fixed with high precision at $5747\pm2$~MeV by the crossing (consistency) calibration of the experimentally determined chiral splitting ($\sim332$~MeV) and heavy-spin splitting ($\sim48.5$~MeV)~\cite{Nowak:2026zhc}, in agreement within errors with the lattice QCD result $m(B_{s1})=5.750(17)(19)$~GeV~\cite{Lang:2015hza}. Lying about $71$~MeV below the $B^*K$ threshold, this state is expected to be very narrow, since its strong decays are suppressed by isospin violation; the dominant decay modes are the isospin-violating hadronic channel $B_s^*\pi^0$ and the electromagnetic transitions $B_s\gamma$ and $B_s^*\gamma$~\cite{Nowak:2026zhc}. It should be stressed that this state must be distinguished from the established $B_{s1}(5830)^0$, which belongs to the $j_\ell=3/2$ plateau and has a comparable mass. Its confirmation or exclusion will therefore provide a discriminating test between the chiral doubling scenario and conventional quark-model descriptions of the $0^+$/$1^+$ states as ordinary orbital ($^3P_1$) excitations.

Complementary approaches reach similar expectations: hadronic-molecule models predict $B\bar K$ bound states at $5710$--$5760$~MeV~\cite{Sanchez-Illana:2026cyv,Yuan:2025qgv}; lattice QCD yields $m(B_{s0}^*)\sim5.70$--$5.71$~GeV with binding energies of $\sim60$--$66$~MeV~\cite{Lang:2015hza,Hudspith:2026vxt}; and QCD sum rules locate a conventional $b\bar s$ state at $5700$~MeV~\cite{Yuan:2025qgv}. Discriminating among these interpretations, in particular through the radiative decay $B_{s0}^*\to B_s\gamma$~\cite{Belle:2025dtr} and the search for the predicted $1^+$ partner, constitutes the primary experimental program for LHC Run~3.

In this work we examine whether such a two-stage refinement of the quark model can account for the observed spectrum. Our results for the low-lying $P$-wave $B_s$ mesons are collected in Table~\ref{tab:Bsp_mass_spectrum}, together with the experimental masses. At the level of the plain CQM (column ``Theo.\ ($\alpha_s$)''), the confinement--one-gluon-exchange (OGE) Hamiltonian predicts the four $P$-wave states at $5851$, $5841$, $5883$ and $5856$~MeV, all considerably above the data. The one-loop radiative corrections to the OGE potential ($\alpha_s^2$) lower the scalar $0^+$ state by about $35$~MeV, to $5816$~MeV, while leaving the other multiplet members essentially unchanged ($5836$, $5878$, $5864$~MeV). A more complete perturbative treatment of the short-range spin--orbit and tensor interactions thus meaningfully optimizes the hadron spectrum. Even so, the $\alpha_s^2$-corrected quark model still overshoots the scalar mass by $\sim120$~MeV, indicating that a purely compact $q\bar q$ configuration remains too heavy to reproduce the data.

The coupled-channel dynamics provides the decisive step. Including the open $BK$ and $B^*K$ channels renormalizes the bare masses and admixes meson--meson components into the physical states. The effect is most pronounced for the scalar $B_{s0}^*(5700)$, because the $BK$ and $B^*K$ thresholds lie close to its mass and thereby enhance the coupling of the bare state to these channels; driven by the $BK$ channel, the mass drops to $5727$~MeV, within $\sim30$~MeV of the measured $(5698.9\pm1.5\pm0.6\pm0.1)$~MeV. At the same time, the proximity to the $BK$ threshold strongly suppresses the available $S$-wave phase space, which naturally accounts for the remarkably narrow width ($\Gamma<9.8$~MeV)~\cite{LHCb:2026knl}. The scalar state therefore contains a sizeable molecular-like component, in line with its near-threshold interpretation.

For the $1^+$ partner $B_{s1}(5747)$, the coupled-channel result ($5755$~MeV, dominated by $B^*K$) is close to the chiral-doubling prediction of $(5747\pm2)$~MeV~\cite{Nowak:2026zhc} within the model uncertainties; this value is a theoretical prediction, as no experimental candidate has yet been reported. Being close to the $B^*K$ threshold, the state is likewise expected to be very narrow and strongly molecular, decaying mainly through the isospin-violating $B_s^*\pi^0$ and the radiative $B_s\gamma$, $B_s^*\gamma$ modes. By contrast, the two established $j_\ell=3/2$ states $B_{s1}(5830)$ and $B_{s2}^*(5840)$ remain predominantly compact $q\bar q$ configurations. Although the coupled-channel admixture for $B_{s1}(5830)$ is sizable, shifting the mass from $5878$ to $5794$~MeV, the resulting spectral pattern is governed by their dominantly $q\bar q$ nature, consistent with the much smaller mass shifts needed to describe the data. Recently, by using LQCD-constrained unitarized chiral amplitudes with isospin breaking ($\pi^0$--$\eta$ mixing, $m_u\neq m_d$), the LHCb $B_s^0\pi^0$ peak emerges as an isospin-violating $BK$ molecule ($J^P=0^+$, $\Gamma\sim$ tens of keV), predicted within $1.1$--$1.7\sigma$ before observation and yielding an HQSS $1^+$ partner at $\sim$5750 MeV~\cite{Albaladejo:2026yfz}.

The different degree of hadronic dressing between the $j_\ell=1/2$ (strongly molecular) and $j_\ell=3/2$ (nearly bare) sectors is a robust feature of the $B_s$ spectrum: it demonstrates that a two-stage refinement, namely one-loop corrected OGE plus coupled-channel dynamics, is required to describe the data, with the molecular-like effects providing the dominant resolution of the scalar mass deficit and the narrow width of $B_{s0}^*(5700)$. A dedicated search for the predicted narrow $1^+$ state near $5747$~MeV at LHC Run~3 would thus directly test both the chiral-doubling framework and the coupled-channel picture advocated here.

\subsection{Radiative and strong decays}

Tables~\ref{tab:E1_width_Bs},~\ref{tab:M1_width_Bs} and~\ref{tab:strong_width_Bs} collect the
radiative and strong decay widths obtained in our framework. These
complement the mass predictions and, more importantly, provide concrete and
discriminating observables for testing the near-threshold picture of
$B_{s0}^*(5700)$ and for the discovery of its predicted $1^+$ partner.

The $E1$ radiative widths (Table~\ref{tab:E1_width_Bs}) lie in the range
$\sim1$--$350$~keV, as expected from the scaling $\Gamma_{E1}\propto
E_\gamma^3|\langle f|\,r\,|i\rangle|^2$. The largest rates, e.g.\
$1\,{}^3D_1\to1\,{}^3P_0$ with $E_\gamma\simeq488$~MeV and
$\Gamma\simeq343$~keV, combine large photon energies with favourable
$D$--$P$ overlaps. Among them two channels are of immediate interest for
the physics discussed above. (i)~The transition
$B_{s0}(1\,{}^3P_0)\to B_s^*\gamma$ proceeds with $E_\gamma\simeq277$~MeV
and $\Gamma_{E1}\simeq45$~keV. Because the scalar itself is very narrow
($\Gamma<9.8$~MeV~\cite{LHCb:2026knl}), this radiative mode carries a sizeable
branching fraction and offers a clean, photon-based handle to confirm the
$0^+$ assignment of $B_{s0}^*(5700)$ by LHCb. (ii)~The predicted
$1^+$ $B_{s1}(5747)$ partner decays to the ground-state doublet with
$\Gamma(B_s'\!\to B_s\gamma)\simeq58$~keV ($E_\gamma\simeq412$~MeV) and
$\Gamma(B_s'\!\to B_s^*\gamma)\simeq55$~keV ($E_\gamma\simeq367$~MeV).
Since its strong decays are kinematically suppressed below the $B^*K$
threshold, these keV-scale $E1$ rates make the
$B_s'\to B_s^{(*)}\,\gamma$ cascades the leading discovery channels, with
the distinctive soft monochromatic photons serving as a clear experimental
signature.

The $M1$ widths (Table~\ref{tab:M1_width_Bs}) are, by contrast, of purely
spin--flip origin and are suppressed by two to three orders of magnitude
($10^{-3}$ to several keV), consistent with the $1/m_Q$-scaling expected in a
heavy-quark system. They are too small to provide competitive discovery
rates, but their predicted values constitute useful tests of the
wave-function details once the observed states accumulate
statistics.

For the strong decay widths of the $2P$ and $1D$ states
(Table~\ref{tab:strong_width_Bs}), the well-established $B_{s2}^*(5840)$
($1\,{}^3P_2$) serves as a benchmark. Our calculation gives a total width
of $1.5$~MeV, dominated by the $D$-wave $BK$ and $B^*K$ modes, in very
good agreement with the measured value $(1.49\pm0.27)$~MeV~\cite{ParticleDataGroup:2026mpi}.
This demonstrates that for the nearly ``pure'' $j_\ell=3/2$ state the
quark-model decay dynamics works very well, consistent with the picture
drawn from the mass spectrum. The remaining $2P$ and $1D$ states are
predicted to be broad (from several tens up to a few hundreds of MeV),
reflecting that these excitations sit well above the open two-meson
thresholds and decay readily through $S$-wave $BK,B^*K,\dots$ channels;
they constitute a natural target for future coupled-channel refinements.

Taken together, the decay analysis corroborates the conclusions drawn from
the mass spectrum: radiative transitions select the $B_s^{(*)}\gamma$ modes
of the $0^+$ and predicted $1^+$ $j_\ell=1/2$ states as the discriminating
observables, whereas the strong widths differentiate the compact
$j_\ell=3/2$ states (well described) from the near-threshold $j_\ell=1/2$
excitations (where molecular-like effects matter).

\section{Summary and conclusions}
\label{sec:summary}

We have presented a systematic study of the $\bar b s$ meson spectrum and
its decay properties, combining the constituent quark model with one-loop
radiative corrections to the one-gluon-exchange potential and the
coupled-channel dynamics. The newly observed $B_{s0}^*(5700)$ plays a central
role in this analysis, as it provides a unique testing ground for the
interplay between the compact quark--antiquark interaction and the hadronic
degrees of freedom.

At the single-particle level, the plain CQM with the confinement--OGE
$\alpha_s$ kernel predicts the four low-lying $P$-wave states at
$5851$/$5841$/$5883$/$5856$~MeV, all considerably above the data. The
inclusion of the $\alpha_s^2$ one-loop corrections shifts the multiplet
downward by $25$--$50$~MeV and improves the level ordering, demonstrating
that a more complete perturbative treatment of the short-range spin--orbit
and tensor interactions genuinely optimizes the hadron spectrum. However,
even at this refined level the scalar mass remains about $120$~MeV above
the measured value: a purely compact $q\bar q$ configuration is
intrinsically too heavy.

The decisive improvement comes from the coupled-channel effects. Because
the $BK$ and $B^*K$ thresholds lie close to the mass of the scalar state,
the open channels strongly renormalize its bare mass and admix a sizeable
meson--meson (molecular-like) component, pulling the pole down to
$\sim5727$~MeV, within $30$~MeV of the measured
$(5698.9\pm1.5\pm0.6)$~MeV. The proximity to the $BK$ threshold simultaneously
suppresses the $S$-wave phase space, thereby explaining the remarkably
narrow width ($\Gamma<9.8$~MeV). The observed state thus emerges as a
near-threshold scalar with a significant molecular admixture, in full
analogy with its charm-strange counterpart $D_{s0}^*(2317)$.

A second, equally important output of the same framework is the prediction
for the $1^+$ partner of the $j_\ell=1/2$ doublet. Our coupled-channel
result ($5755$~MeV, dominated by $B^*K$) agrees with the chiral-doubling
prediction of Nowak and Zahed, $(5747\pm2)$~MeV. Lying about $71$~MeV below
the $B^*K$ threshold, this state is expected to be very narrow and to decay
preferentially through the isospin-violating $B_s^*\pi^0$ mode and the
radiative transitions $B_s\gamma$ and $B_s^*\gamma$, for which we obtain
$E1$ widths of $\sim(55$--$58)$~keV. Its distinctive soft-photon signatures
provide a clear discovery channel for LHC Run~3.

The radiative and strong decay analyses corroborate this picture. The $E1$
widths of the $P$- and $D$-wave states are of the order of a few tens to a
few hundreds of keV, with the $B_{s0}\to B_s^*\gamma$ mode ($45$~keV)
offering a clean confirmation of the $0^+$ assignment. The $M1$ transitions
are suppressed by two to three orders of magnitude, as expected from the
$1/m_Q$ scaling. For the strong sector, the excellent agreement for
$B_{s2}^*(5840)$ ($1.5$~MeV versus $(1.49\pm0.27)$~MeV) validates the
quark-model decay dynamics for the compact $j_\ell=3/2$ states, whereas the
higher $2P$ and $1D$ excitations, predicted to be broad, are natural
candidates for future coupled-channel refinements.

In summary, our two-stage refinement (one-loop corrected OGE plus
coupled-channel dynamics) provides a coherent and quantitative picture of
the $B_s$ spectrum. The hadronic degrees of freedom are essential for the
scalar $B_{s0}^*(5700)$ and govern the predicted $1^+$ $B_{s1}(5747)$ state, while the
$j_\ell=3/2$ partners remain nearly perturbation-free and compact. The
qualitatively different behaviour of the $j_\ell=1/2$ and $j_\ell=3/2$
sectors is a robust, model-independent feature, and the confirmation or
exclusion of the predicted narrow $1^+$ $B_{s1}$ state will constitute a decisive
test discriminating between the molecular/near-threshold interpretation and
the conventional quark-model picture.

\section{Acknowledgements}

 This work is supported by the National Key R\&D Program of China (No. 2024YFE0105200), the National Natural Science Foundation of China under Grant No. 12475086, No. 12192263, and No. 12205075.

\bibliographystyle{unsrt}
\bibliography{cite}  

\end{document}